\documentstyle[12pt,epsfig]{article}
 1
\def\citenum#1{{\def\@cite##1##2{##1}\cite{#1}}}
\def\citea#1{\@cite{#1}{}}

\def\b{\beta}
\def\a{\alpha}

\def\l{\lambda}

\def\O{\Omega}

\def\s{\sigma}

\def\({\left(}   
\def\){\right)}

\def\citenum#1{{\def\@cite##1##2{##1}\cite{#1}}}
\def\citea#1{\@cite{#1}{}}

\def\l1vt{\vec{l_{1\perp}}}

\def\bt{b_{\perp}}

\def\bt2{$b^2_t$}

\def\jol1{$J_0(\,l_{1\perp}\,r_{\perp}\,)$}

\def\citea#1{\@cite{#1}{}}

\def\beq{\begin{equation}}
\def\eeq{\end{equation}}
\def\bea{\begin{eqnarray}}
\def\eea{\end{eqnarray}}

\def\eq#1{{Eq.~(\ref{#1})}}

\def\bbbz{{\mathchoice {\hbox{$\sf\textstyle Z\kern-0.4em Z$}}
{\hbox{$\sf\textstyle Z\kern-0.4em Z$}}
{\hbox{$\sf\scriptstyle Z\kern-0.3em Z$}}
{\hbox{$\sf\scriptscriptstyle Z\kern-0.2em Z$}}}}

\def\npb#1#2#3{    {\it Nucl. Phys. }{\bf B#1} (19#2) #3}
\def\plb#1#2#3{    {\it Phys. Lett. }{\bf B#1} (19#2) #3}
\def\prd#1#2#3{    {\it Phys. Rev. }{\bf D#1} (19#2) #3}

\def\prl#1#2#3{    {\it Phys. Rev. Lett. }{\bf #1} (19#2) #3}

\relax
\begin{document}
\newcounter{savefig}
\newcommand{\alphfig}{\addtocounter{figure}{1}%
\setcounter{savefig}{\value{figure}}%
\setcounter{figure}{0}%
\renewcommand{\thefigure}{\mbox{\arabic{savefig}-\alph{figure}}}}
\newcommand{\resetfig}{\setcounter{figure}{\value{savefig}}%
\renewcommand{\thefigure}{\arabic{figure}}}

\begin{titlepage}
\begin{flushright}
TAUP\,\,2560 - 99\\
 \today\\
\end{flushright}

\centerline{}

\centerline{}

\centerline{}

\begin{center}
{\Large\bf{A TWO CHANNEL  CALCULATION}}\\[1.5ex]
{\Large \bf{OF SCREENING CORRECTIONS}}\\[6ex]

{\large \bf { E. Gotsman ${}^*$\footnotetext{ ${}^*$ E-mail:
gotsman@post.tau.ac.il}
,
 E. Levin $^{\star}$\footnotetext{${}^{\star}$
E-mail:leving@post.tau.ac.il}
  and 
 U. Maor ${}^{\dagger}$
\footnotetext{$^{\dagger}$ E-mail: maor@post.tau.ac.il}}} \\[2.5ex]

{\it School of Physics and Astronomy, Tel Aviv University}\\
{\it Ramat Aviv, 69978, ISRAEL}\\[10.5ex]

{\large \bf Abstract:}

\end{center}

We present a two channel eikonal calculation in which the
rescattering through a diffractive channel is included in addition to the 
elastic channel. Considering the spread of the experimental data,  
we find that we can obtain a very good  description of
 $\s_{tot}$, $\s_{el}$ and  $B_{el}$ in the ISR -
Tevatron energy range. In this range of energy the  diffractive channel,
that was included in our
calculation, leads to a ratio  of  
$\s_{SD}/\s_{el}$ which varies between 1 and
0.5 for $ 20\,GeV\,\,\leq\,\sqrt{s}\,\leq\,14\,TeV$   in
  agreement with the experimental data.
 The calculated survival probability of dijet
production  with a large rapidity gap is consistent with the data. 

\end{titlepage}

\section{Introduction}
A simple Regge-pole parameterization suggested by Donnachie and
Landshoff (DL)\cite{DL}, has achieved remarkable success in describing
hadronic and photon induced total and elastic cross sections. The
elastic amplitude is, obviously, bounded by s-channel unitarity
requiring screening (shadowing) corrections (SC) at sufficiently high
energies. However, up   to the Tevatron
energies  no decisive experimental signature of
this effect
has, thus far, been seen  in the elastic channel.

An overall view of the features of high energy  collisions
\cite{DATA}\cite{SD}\cite{D0}\cite{CDF}\cite{ZEUS}\cite{H1}   
reveals at least two channels which are not compatible with the DL
parameterization.

\begin{enumerate}

\item\,\,\, Since the total cross section behaves at high energies as
$s^{\Delta_P}$,
where $\alpha_{P}(0) = 1 + \Delta_P$, we expect
both $\sigma_{el}$ and $\sigma_{SD}$, the integrated single diffraction
(SD) cross section, to behave approximately like $s^{2\Delta_P}$ (since
the
Pomeron trajectory slope is small we can neglect the effects of the t
integration). This is, indeed, the experimentally observed behaviour of
$\sigma_{el}$, but $\sigma_{SD}$ has a much milder energy dependence
clearly seen in the ISR-Tevatron energy range\cite{SD}.

\item\,\,\, $< |S|^2 >$, which is
defined\cite{BJ} as the survival probability of two
jets with a large rapidity gap (LRG), is measured\cite{D0}\cite{CDF} to
be a small fraction of unity decreasing with energy. In a DL type model, 
containing very weak rescattering corrections, we expect
$< |S|^2 >$ to be approximately one with no, or very little,
dependence on energy.

\end{enumerate}

These problems have been discussed in our previous
publications\cite{ZEIT}\cite{S2}\cite{PRD}\cite{2P}\cite{S2N} where
we have utilized a simplified one channel eikonal approximation in
which the b-space scattering amplitude is given by
\beq\label{1.1}
a(s,b)\,\,\,=\,\,\,i\,\left(\,1\,\,-\,\,e^{-
\,\frac{\O(s,b)}{2}}\,\right)\,\,,
\eeq
 the opacity is assumed to have a Gaussian form
\beq\label{1.2}
\O(s,b)\,\,\,=\,\,\,\nu(s)\,e^{- \,\frac{b^2}{R^2(s)}}\,\,.
\eeq

Even though this may be considered as an over simplified toy model, it
provides a semi realistic reproduction of the main
experimental features.
Our main conclusion\cite{PRD}\cite{2P} is that
while SC saturate the elastic channel in the TeV range,
diffractive channels are saturated
at relatively low (ISR) energies.
This feature results in a mild energy dependence of $\sigma_{SD}$
at fixed $M^2$, where
the power behaviour of the cross section with $s$ approaches a $ln s$
dependence in the high energy limit.
The above eikonal approximation also
leads\cite{S2}\cite{S2N} to estimates of
$< |S|^2 >$ which are compatible
with the Tevatron\cite{D0} and HERA\cite{ZEUS} data.   

The one channel eikonal approach contains
two severe deficiences::
\begin{enumerate}

\item\,\,\, The main input assumption in such a model is that
the rescattering is elastic, i.e. $\sigma_{diff} << \sigma_{el}$.
This is clearly not corroborated by the data in the ISR-Tevatron energy
range.

\item\,\,\, The introduction of SC introduces a
fundamental problem in the
definition of $\sigma_{SD}$. In a missing mass experiment,
the integrated   
SD cross section is defined as
\beq\label{1.3}
\s_{SD}\,\,=\,\,\int^{t_{max}(M^2)}_{t_{min}(M^2)}\,
\int^{M^2_{max}}_{M^2_{min}}\,\,\,\frac{d^2\,\s_{SD}}{d M^2 \,d t}\,\,,
\eeq  
where the upper $M^2$ integration limit is taken as a fixed
fraction of $s$, usually 0.05$s$.
In a non screened triple Regge approximation \cite{MU}  
$\sigma_{SD}$ behaves like $s^{2\alpha_P(t)-2}$
at fixed $M^2$. With a super critical Pomeron ($\alpha_p\,>\,1$),
the $M^2$ integration converges,
leading to an approximate $s^{2\Delta_P}$
behaviour of the integrated cross section.
However, once SC are introduced\cite{PRD}, the $M^2$
dependence of $\sigma_{SD}$ changes
in the asymptotic limit to $ln s$.
Accordingly, the $M^2$ integration is divergent in
$s$ due to the upper $M^2$ integration limit. The  predictions are, thus,
flawed if examined over a wide energy range.   
Indeed, while we managed  to fit
$\sigma_{SD}$ in the
UA(4)-Tevatron range, our calculated $\sigma_{SD}$ values \cite{PRD} in
the ISR
energy range fall below the data \cite{SD}.

\end{enumerate}

\section{The Model}
Our goal is to construct an eikonal model in which the rescattering can be
either elastic or diffractive. In the simplest approximation we consider 
the diffractively produced hadrons as a single hadronic state. We have, 
thus, two orthogonal wave functions
\beq\label{2.1}
< \Psi_h | \Psi_D >\,\,=\,\,0
\eeq
where $\Psi_h$ is the wave function of the incoming hadron, and $\Psi_D$ 
is the wave function of the outgoing  diffractively produced particles,
initiated by this hadron.

Consider   two wave functions $\Psi_1$ and $\Psi_2$ which are 
diagonal with respect to $\large\bf{T}$, the interaction operator.
The amplitude  of the high energy interaction is
equal to
\beq \label{2.2}
A_{i,k}\,\,=\,\,< \Psi_{i} \Psi_{k} |{ \large \mathbf T} | \Psi_{i'}
\Psi_{k'}
>\,\,=\,\,A_{i,k}\,\delta_{i,i'}\,\delta_{k,k'}\,\,.
\eeq
In a two channel model $i,k  = 1,2$. The extension to a higher number 
of channels is straight forward. The amplitudes $A_{i,k}$ satisfy the
diagonal 
unitarity condition 
\beq\label{2.3}
2\,\,Im A^{el}_{i,k} (s, b )\,\,=\,\,| A^{el}_{i,k} (s, b )|^2\,\,
+\,\,G^{in}_{i,k} (s, b)\,
\eeq
for which we write the solution
\bea
&
A^{el}_{i,k}(s, b )\,\,\,=\,\,\,i\,\left( \,1\,\,\,-\,\,\,e^{ -
\,\frac{\O_{i,k}(s,b)}{2}}\,\right)\,\,;&\label{2.4}\\
&
G^{in}_{i,k} (s, b_t)\,\,\,=\,\,\,1\,\,\,-\,\,\,e^{ -
\,\O_{i,k}(s,b)}\,\,,&\label{2.5}
\eea
where $\Omega_{i,k}$ is the opacity of the $(i,k)$-th channel with a
wave function $ \Psi_i\,\times \,\Psi_k$.

In this representation $\Psi_h$ and $\Psi_D$ can be written as
\bea 
&
\Psi_{h}\,\,\,=\,\,\a\,\Psi_1\,\,\,+\,\,\,\,\b\,\Psi_2\,\,;&\label{2.6}\\
&
\Psi_{D}\,\,\,=\,\,-\,\b\,\Psi_1\,\,\,+\,\,\,\,\a\,\Psi_2\,\,\;\label{2.7}&
\eea
Since $|\Psi_h|^2 = 1$, we have
\beq\label{2.8}
\a^2\,\,\,+\,\,\b^2\,\,\,=\,\,\,1
\eeq
The wave function of the final state is 
\beq\label{2.9}
\Psi_f\,\,=\,\,| T | \Psi_h \,\times\,\,\Psi_h >\,\,\,=
\eeq
$$
\a^2\,A_{1,1}\,\Psi_1 \,\times\,\Psi_1\,\,+\,\,\a\,\b\,A_{1,2}\,\{\,\Psi_1
\,\times\,\Psi_2 \,+\,\Psi_2
\,\times\,\Psi_1\,\}\,\,+\,\,\b^2\,A_{2,2}\,\,\Psi_2
\,\times\,\Psi_2
$$
We   now  define
\bea
&
a_{el} (s, b)\,\,=\,\,<\, \Psi_{h}\,\times\,
\Psi_{h} | \Psi_{f} \,>\,\,=\,\,
\a^4\,A_{1,1}
\,\,+\,\,2\,\a^2\,\b^2\,A_{1,2}\,\,+\,\,\b^4\,A_{2,2}\,;&\label{2.101}\\
&
a_{SD} (s, b)\,\,=\,\,<\, \Psi_{h}\,\times\,\Psi_D | \Psi_{f} \,>\,\,=
\,\,\a\,\b\,\{\,\,-\,\a^2\,A_{1,1} \,\,+\,\,(\,\a^2\,-\,\b^2\,)
\,A_{1,2}\,\,+\,\,\b^2\,\,A_{2,2}\,\,\}\,\,;&\label{2.102}\\
&
a_{DD} (s, b)\,\,=\,\,<\, \Psi_D\,\times\,\Psi_D | \Psi_{f} \,>\,\,=
\a^2\,\b^2\,\{\,\,A_{1,1} \,\,-\,\,2
\,A_{1,2}\,\,+\,\,A_{2,2}\,\,\}\,\,.&\label{2.103}
\eea
Assuming, that the double diffractive production is small, we  have  
\bea
&
a_{el}(s,b)\,\,=\,\,A_{1,1}
\,\,-\,\,2\,\b^2\,\left(\,A_{1,1}\,\,-\,\,A_{1,2}\,\right)\,\,;&\label{2.10}\\
&
a_{SD}(s,b)\,\,=\,\,-\,\a\,\b\,\left(\,A_{1,1}\,\,-\,\,A_{1,2}\,\right)\,\,.&
\label{2.11}
\eea

 $a_{el}$ and $a_{SD}$ can be written in terms of the opacities
$\Omega_1\,\equiv\,\O_{1,1}$ 
and $\Omega_2\,\equiv\,\O_{1,2}$
\bea
&
a_{el}(s,b)\,\,=\,\,i\,\left(\,1 \,-\,e^{- \frac{\O_1(s,b)}{2}}\,)\right)
\,\,-\,\,2\,\b^2\,\left(\,e^{- 
\frac{\O_1(s,b)}{2}}\,\,-\,\,e^{-\frac{\O_2(s,b)}{2}}\,\right)\,\,;&
\label{2.12}\\ &
a_{SD}(s,b)\,\,=\,\,-\,\a\,\b\,\left(\,e^{-\frac{\O_1(s,b)}{2}}\,\,-
\,\,e^{- \frac{\O_2(s,b)}{2}}\,\right)\,\,.&
\label{2.13}
\eea
 Defining $\Delta \Omega = \Omega_2 - \Omega_1$, we get
\bea
&
a_{el}(s,b)\,\,=\,\,i\,\left(\,1 \,-\,e^{- \frac{\O_1(s,b)}{2}}\,)\right)
\,\,-\,\,2\,\b^2\,e^{-\frac{\O_1(s,b)}{2}}\,\left(\,
1\,\,-\,\,e^{-\frac{\Delta \O(s,b)}{2}}\,\right)\,\,;&
\label{2.14}\\ &
a_{SD}(s,b)\,\,=\,\,-\,\a\,\b\,e^{-\frac{\O_1(s,b)}{2}}\,\left(\,1\,\,-
\,\,e^{- \frac{\Delta\,\O(s,b)}{2}}\,\right)\,\,.&
\label{2.15}
\eea
In the limit where  $\beta << 1$ and $\Delta \Omega << 1$, we reproduce 
the single channel eikonal model
\bea
&
a_{el}(s,b)\,\,=\,\,i\,\left(\,1\,\,-\,\,e^{- \frac{\O_1(s,b)}{2}}\,)\right)
\,\,;&\label{2.16}\\
&
a_{SD}(s,b)\,\,=\,\,\b\,\frac{\Delta \,\O(s,b)}{2}\,e^{-
\frac{\O_1(s,b)}{2}}\,\,.&\label{2.17}
\eea

For a convenient semi realistic parameterization, we follow our earlier
publications  and adopt a Gaussian approximation for 
the opacities
\beq\label{2.18}
\O^P_1(s,b)\,\,=\,\,\frac{\sigma^P_{01}}{\pi ( R^P_1(s))^2}\,(
\frac{s}{s_0})^{\Delta_P}\,e^{-\frac{b^2}{(R^P_1(s))^2}}\,\,=\,\,\nu^{0P}_1
(\frac{s}{s_0})^{\Delta_P}\,e^{-\frac{b^2}{(R^P_1(s))^2}}\,;
\eeq

\beq\label{2.19}
\Delta\O^P(s,b)\,\,=\,\,\frac{\sigma^P_{0D}}{\pi ( R^P_D(s))^2}\,(
\frac{s}{s_0})^{\Delta_P}\,e^{-\frac{b^2}{(R^P_D(s))^2}}\,\,=\,\,\nu^{0P}_D
(\frac{s}{s_0})^{\Delta_P}\,e^{-\frac{b^2}{(R^P_D(s))^2}}\,.
\eeq
For both radii we assume the form  
\beq\label{2.20}
( R^P_i )^2\,\,=\,\,(  R^P_{0i}
)^2\,\,\,+\,\,\,4\,\alpha'_P\,\,\ln(s/s_0)\,\,,
\eeq
where $i\,\,=\,\,1, D$, and take  $s_0\, = \,1 \,GeV^2 $.

Since our investigation covers the ISR energy range , we need to include,
in 
addition to the Pomeron parameters defined above, a Regge contribution 
which is defined in a similar way
\beq\label{2.21}
\O^R_1(s,b)\,\,=\,\,\frac{\sigma^R_{01}}{\pi ( R^R_1(s))^2}\,(
\frac{s}{s_0})^{-\eta}\,e^{-\frac{b^2}{(R^R_1(s))^2}}\,\,=\,\,\nu^{0R}_1
(\frac{s}{s_0})^{- \eta}\,e^{-\frac{b^2}{(R^R_1(s))^2}}\,\,;
\eeq
\beq\label{2.22}
\Delta\O^R(s,b)\,\,=\,\,\frac{\sigma^R_{0D}}{\pi ( R^R_D(s))^2}\,(
\frac{s}{s_0})^{-
\eta}\,e^{-\frac{b^2}{(R^R_D(s))^2}}\,\,=\,\,\nu^{0R}_D
(\frac{s}{s_0})^{- \eta}\,e^{-\frac{b^2}{(R^R_D(s))^2}}\,\,;
\eeq
\beq\label{2.23}
( R^R_i
)^2\,\,=\,\,(R^R_{0i})^2\,\,\,+\,\,\,4\,\alpha'_R\,\,\ln(s/s_0)\,\,;
\eeq
where $\eta \,\,=\,\,1 \,\,-\,\,\a_R(0)$.

In the following we check the ability of this simple model to describe
the high energy data on
\bea
&
\s_{tot}( s )\,\,=\,\,2\,\pi\,\int^{\infty}_0\, d b^2
\,\,a_{el}(s,b)\,\,;&\label{2.24}\\
&
\s_{el}( s )\,\,=\,\,\pi\,\int^{\infty}_0 \,d
b^2\,\,|\,a_{el}(s,b)\,|^2\,\,;&
\label{2.25}\\
&
B_{el}( s )\,\,=\,\,\frac{\int^{\infty}_0 d
b^2\,\,b^2\,\,a_{el}(s,b)}{2\,\,
\int^{\infty}_0 \,d b^2\,\,a_{el}(s,b)}\,\,;&\label{2.26}\\
&
\s_{SD}( s )\,\,=\,\,2\,\pi\,\int^{\infty}_0 d b^2\,|\,a_{SD}(s,b)\,|^2
\,\,;&\label{2.27}
\eea
where we use $\Omega_i(s,b)\,\,=\,\,\O^P_i(s,b)\,\,+\,\,\O^R_i(s,b)$.

We assume single diffraction to be the only non negligible diffractive
channel.  By averaging  the $pp$ and $p \bar p$ data, we eliminate
the need to discuss odd parity Regge
exchanges, which may be marginally important at the lower ISR energies.
 The elimination of the
odd contribution is exact for $\sigma_{tot}$ and approximate for the
elastic channel.

Once the parameters of our model are specified we can calculate the
survival probability \cite{BJ} of dijets with a LRG. We consider this is
an important consistency check and discuss it in Section 4.

\section{Data base and fitting procedures}
Our study aims at a better understanding of the role s-channel unitarity 
screening and its relevance to soft Pomeron physics. Accordingly, we have 
limited our investigation to high energy $pp$ and $p \bar p $ data at the 
ISR and above, i.e. $s > 300\, GeV^2$. This choice is made so as to
minimize
the dependence of our analysis on the secondary Regge exchanges.

Our data base contains 61 entries:
\begin{enumerate}

\item\,\,\, 18 values for $\sigma_{tot}$ \cite{DATA}. In the ISR range we
have 10
points
averaged between $pp$ and $p \bar p$. Above ISR there are 8 measured
values
of $\sigma_{tot}(p \bar p)$. These energies are sufficiently high to
neglect the odd parity contributions.

\item\,\,\, 9 \, $\sigma_{el}$ data points\cite{DATA}, 5 of which are
averaged ISR
values and 4
higher energy cross sections.

\item\,\,\, 11 values for $B_{el}$\cite{DATA}. For 2 ISR energies, where
we have the
data, we 
took the averaged value.

\item\,\,\, 21 $\sigma_{SD}$ data points\cite{SD}, all of which were
obtained from
missing 
mass experiments initiated by $p \bar p$. The data corresponds to the sum       
of SD produced at either the proton or anti proton vertex. We have omitted
from the analysis the SD data reported by Ref.\cite{SDU}. These data
points are systematically much lower than the other ISR reported SD cross 
sections\cite{SD}.
\end{enumerate}

Our model, as specified in Section 2, contains Pomeron and effective Regge
trajectories
\beq\label{3.1}
\alpha_P(t)\,\,\,=\,\,\,1\,+\,\Delta_P\,\,+\,\,0.2\,t\,\,;
\eeq
\beq\label{3.2}
\alpha_R(t)\,\,\,=\,\,\,0.6\,\,\,+\,\,\,t\,\,;
\eeq
where $\Delta_P$ is a fitted parameter. The assumed  values of
$\alpha_p^{,}$, 
$\alpha_R(0)$ and $\alpha_R^{,}$ are those  commonly used in Regge
phenomenology. 
The other fitted parameters, defined in Section 2,  
are $\beta$, $\Delta_P$, $\sigma_{01}^P$, $( R_{01}^P )^2$, 
$\sigma_{0D}^P$, $\sigma_{01}^R$, $( R_{01}^R )^2$ and $\sigma_{0D}^R$.
In accordance with the triple Regge formalism we take 
\beq\label{3.3}   
(\,R^i_{0D}\,)^2\,\,\,=\,\,\frac{1}{2}\,(\,R^i_{01}\,)^2\,\,\,\,\,\,\,{\rm
where}\,\,\,\,\,\,i\,\,=\,\,P\,,\,\,R\,\,,
\eeq
We neglect the triple vertex radius\cite{PRD}.
  
\begin{figure}
\centerline{\epsfig{file=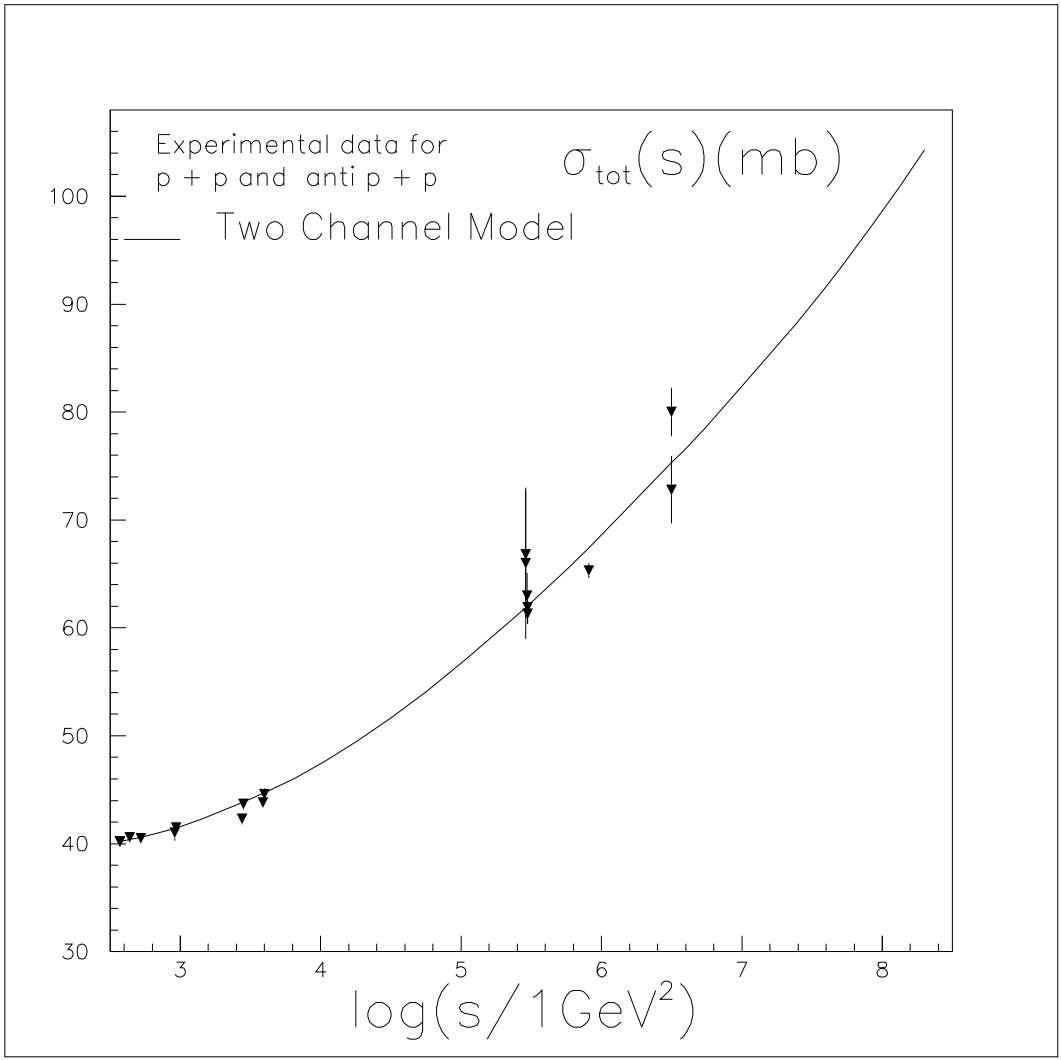,width=140mm}}
\caption{\it Total $p p $ and $\bar p p $ cross sections versus energy. 
}
\label{Fig.1}
\end{figure}

\begin{figure}
\centerline{\epsfig{file=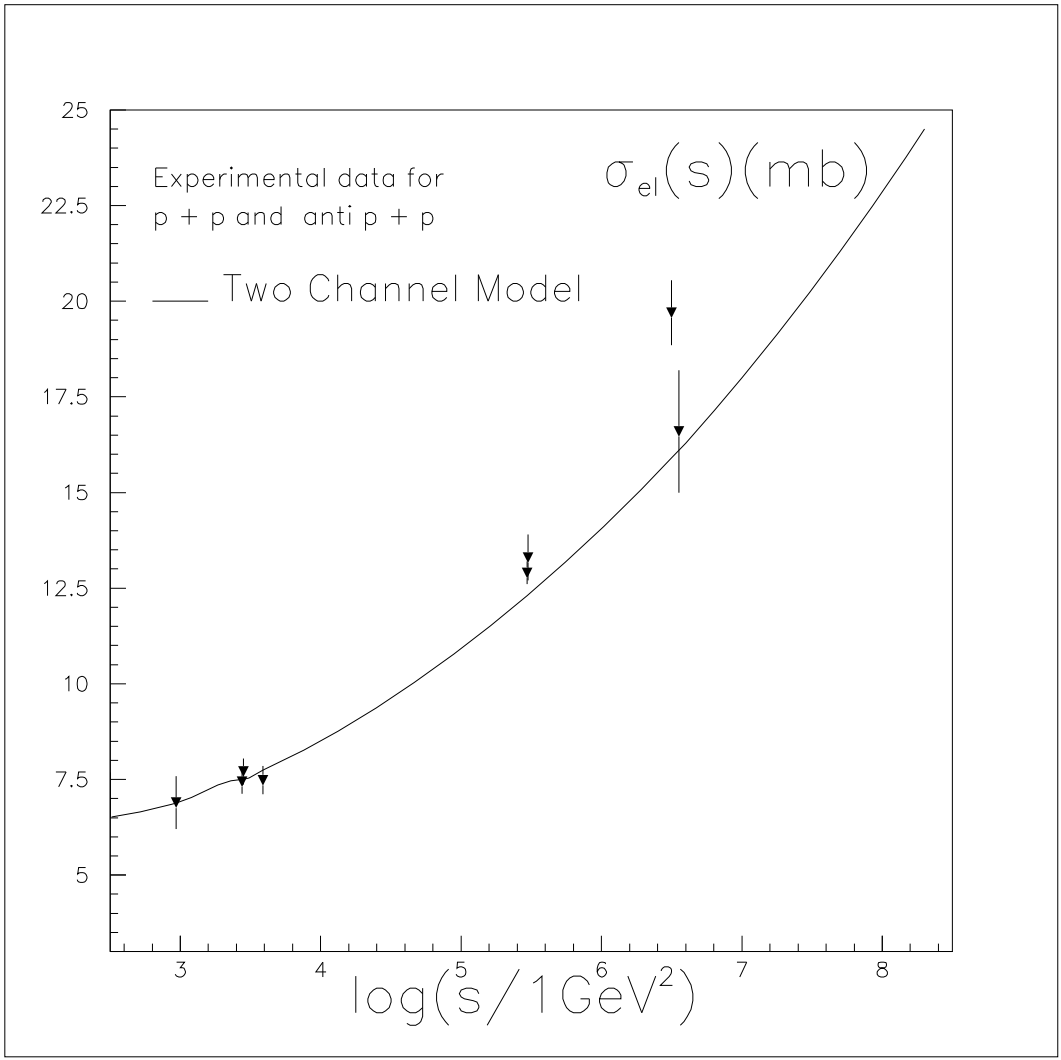,width=140mm}}
\caption{\it Elastic cross section versus energy. }
\label{Fig.2}
\end{figure}

\begin{figure}
\centerline{\epsfig{file=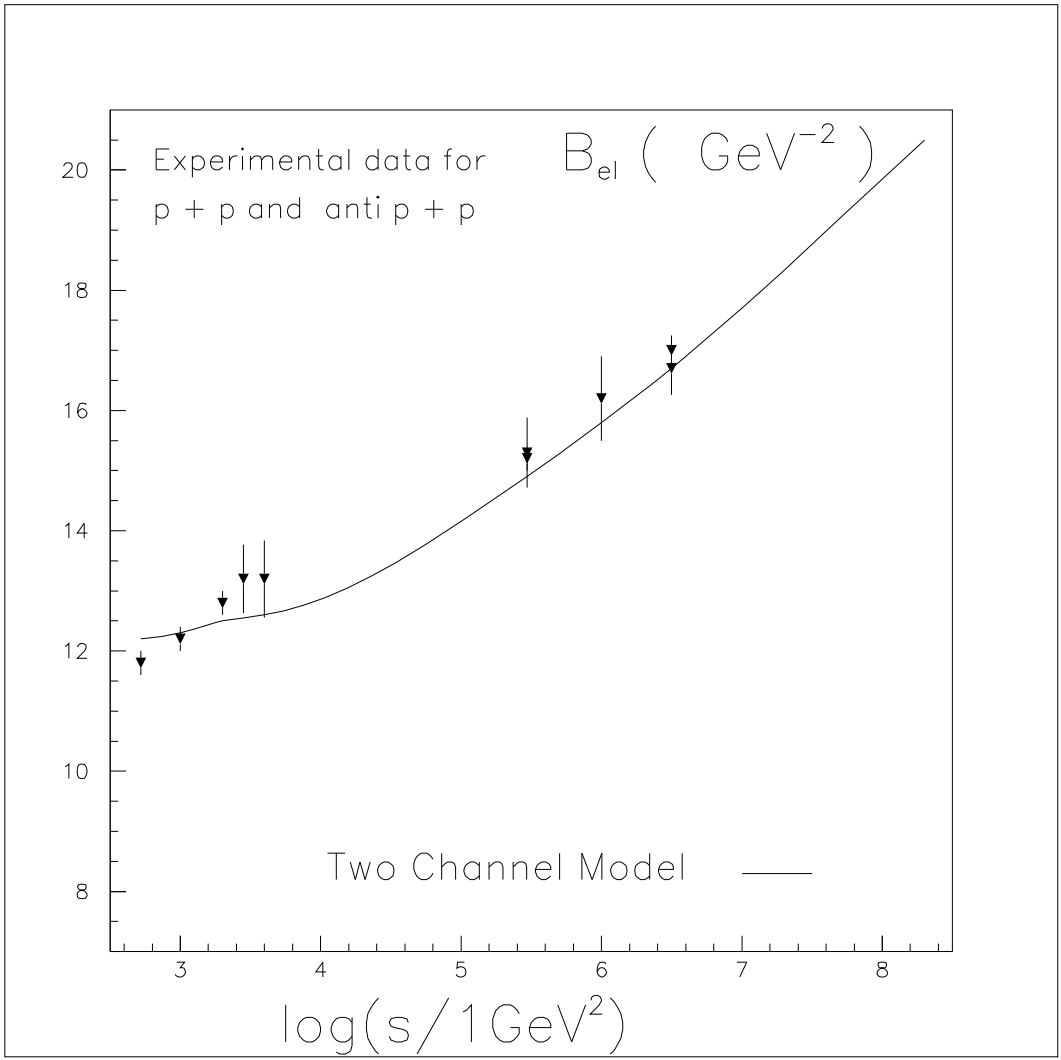,width=140mm}}
\caption{\it Slope of the elastic cross section versus energy. }
\label{Fig.3}
\end{figure}

\begin{figure}
\centerline{\epsfig{file=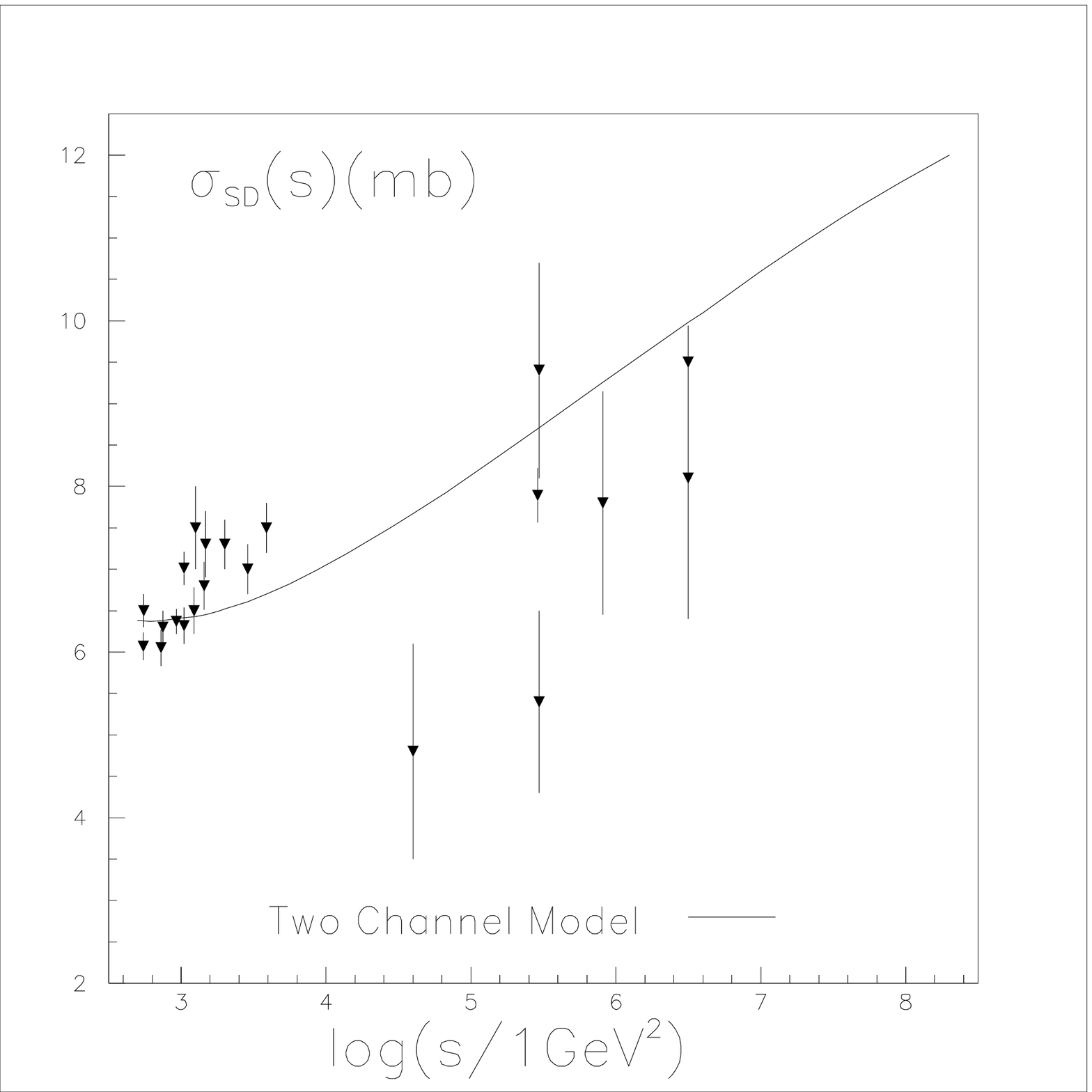,width=140mm}}
\caption{\it Single diffraction cross section  versus energy. }
\label{Fig.4}
\end{figure}

The predictions of our best fit,  compared with the data are presented
in Figs 1-4. The values of the fitted parameters are given
in Table I. Our best fit has an overall $\chi^2 = 129.8 $ corresponding to 
$\frac{\chi^2}{d.f.} = 2.6$. This is  a non satisfactory high
value. However, when inspecting the $\chi^2$ contribution of the
individual points we find  that 7 points produce $\chi^2= 65.2$. 
Neglecting these 7 points results in $\frac{\chi^2}{d.f.} = 1.5$.
Re-fitting the data without these 7 points, we get essentially the same
fit. Conidering the spread of the experimental points reported with small
quoted errors, we consider our fit to be a  good one.

\begin{table}
\parbox{7cm}{
\begin{flushleft}

\begin{tabular}{|l | l |}
\hline
   &  \\ 
Parameters & Best fit values\\
 & \\ \hline
   & \\
$\Delta_P$ & 0.126 \\ 
  & \\
   &  \\ \hline
  & \\
$\beta$ & 0.464 \\ 
   &  \\ \hline
  & \\
$\sigma^P_{01} \,(\,GeV^{-2} \,)$ & 12.99 \\ 
   &  \\ \hline
  & \\
$(\,R^P_{01}\,)^2\,(\,GeV^{-2}\,)$ &   16.34 \\ 
   &  \\ \hline
  & \\
$\sigma^P_{0D} \,(\,GeV^{-2} \,)$ & 145.6 \\ 
   &  \\ \hline
  & \\
$\sigma^R_{01} \,(\,GeV^{-2} \,)$ & 4.78 \\ 
   &  \\ \hline
  & \\
$(\,R^R_{01}\,)^2\,(\,GeV^{-2}\,)$ &  12.44 \\ 
   &  \\ \hline
  & \\
$\sigma^R_{0D} \,(\,GeV^{-2} \,)$ & 999.0\\ 
   &  \\ \hline
\end{tabular}
\end{flushleft}}
\parbox{6cm}{
\begin{flushleft}
\caption{\it The values of the parameters of the Pomeron and Reggeon input
Eikonal amplitudes.}
\end{flushleft}}
\end{table}

\section{Survival probabilities}

A LRG process is identified by the absence of produced hadrons in a 
sufficiently large rapidity gap region. This is regarded as a reliable 
signature for a crossed channel exchange of a Pomeron or alternatively a 
colourless gluonic state, be it non perturbative or perturbative. Recent 
experimental investigations of jets produced in the
Tevatron\cite{D0}\cite{CDF} and HERA\cite{ZEUS}\cite{H1} have recorded
this phenomena. We shall discuss here the simplest process where we have 
two jets produced with a LRG separation.

\begin{figure}
\centerline{\epsfig{file=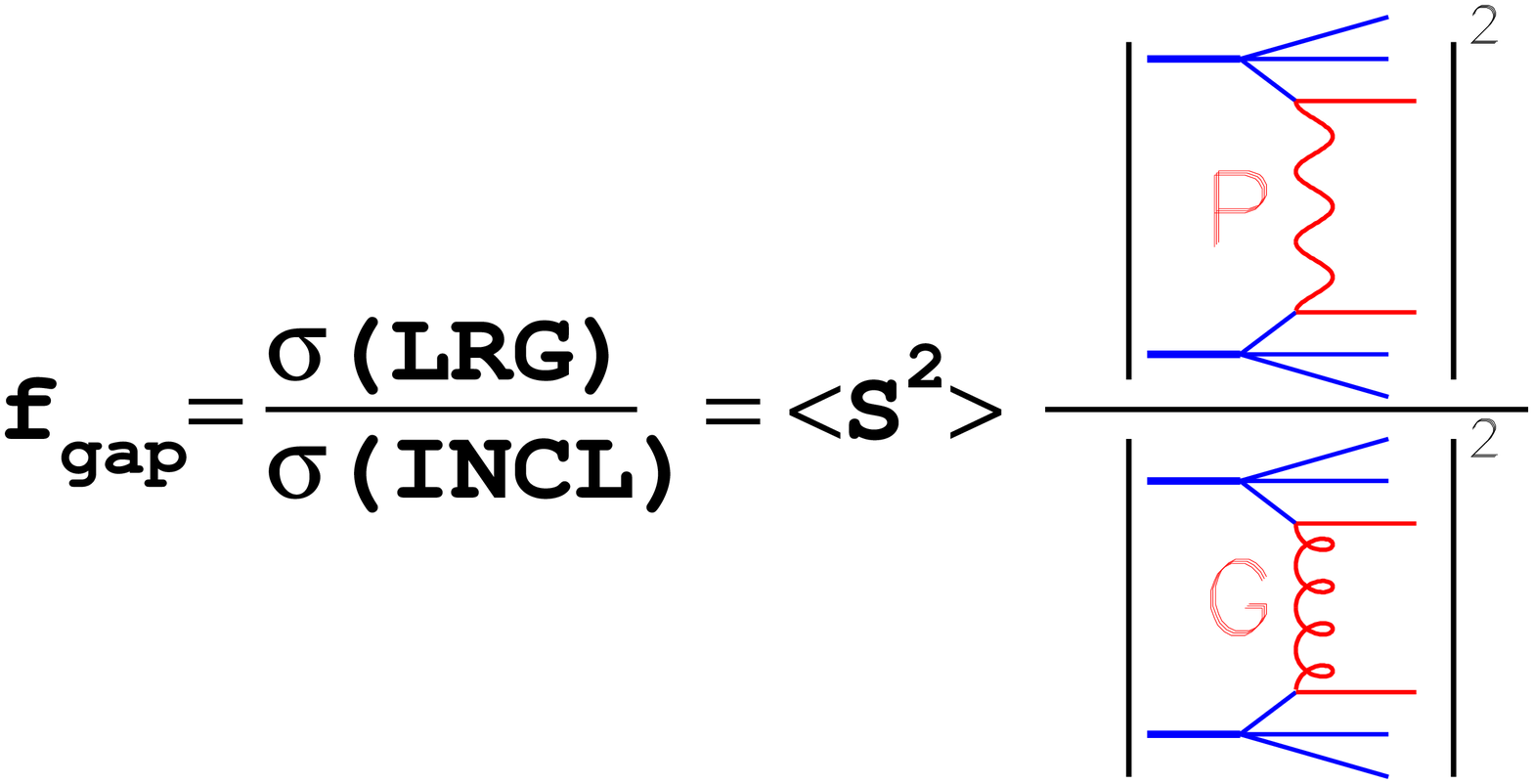,width=140mm}}
\caption{\it Pictorial definition of $f_{gap}$, where P and G represent,
respectively, the exchange of a colour singlet and a colour octet. }
\label{Fig.5}
\end{figure}

The experimentally measured quantity of interest is $f_{gap}$, which is
the 
ratio of the cross section of dijet production with a LRG,  
and the inclusive two jet cross section (see Fig. 5)
\beq\label{4.1}
f_{gap}\,\,\,=\,\,\,\frac{\s_{LRG}( 2 \,\,\,jets )}{\s_{incl}( 2
\,\,\,jets )}\,\,\,=\,\,\,<\,\mid\,S\,\mid^2\,>\,\cdot\,F_s\,\,,
\eeq
$F_s$ is the dynamical ratio of singlet to octet colour  exchange leading
to two
jet production and its calculation is beyond the scope of this paper. 
$F_s$ has to be modified by the survival probability $< |S|^2 >$, which 
is the probability that the produced LRG event is not filled by hadronic 
debris resulting from rescattering of partons or hadrons.

The calculation of $< |S|^2 >$ is model dependent. We shall follow the
procedure suggested by Bjorken\cite{BJ}, in which the survival probability
is the normalized b-space convolution of the hard partonic scattering
amplitude and $P(s,b)$ - the probability that the two initial projectiles
do not interact inelastically. 
\beq\label{4.2}
< \mid S(\,s\,) \mid^2>\,\,\,=\,\,\,\frac{\int\,\,d^2 b
\,P(s,b)\,a_{H}(\Delta y,b )}{\int\,\,d^2 b a_{H}(\Delta y,b)}
\,\,.
\eeq
In the following  we neglect the dependence of $a_H$ on the rapidity gap
of
interest.

In a single channel eikonal model
\beq\label{4.3}
P(s,b)\,\,\,=\,\,\,e^{ - \O(s,b)}\,\,.
\eeq
Taking a Gaussian approximation for both $\Omega(s,b)$ and $a_H(s,b)$ 
simplifies the calculation and we obtain
\beq\label{4.4}
< \mid S (\,s\,) \mid^2> =  \frac{a \gamma[a,\nu]}{\nu^{a}}\,\,,
\eeq
where $\gamma(a,x) = \int_0^x z^{a-1} e^{-z} dz$, $\nu$ is defined in
\eq{1.2} and $a = \frac{R_s^2(s)}{R_H^2(s)}$.

 The results  of our single channel eikonal calculation\cite{S2} are  in
 fair agreement
 with  the D0 data\cite{D0}. However, the calculated energy
dependence is not sufficiently strong. This deficiency is removed once we 
estimate\cite{S2N} $\nu, R_s^2 $ and $ R_H^2$ directly from the data.
The two channel eikonal model presented in this paper may have
some problems in reproducing the  experimental energy  dependence, since
 the data for  
 $\frac{\sigma_{el} + \sigma_{diff}}{\sigma_{tot}}$ is
 almost energy independent in the ISR-Tevatron energy range,
whereas $\frac{\sigma_{el}}{\sigma_{tot}}$  grows  monotonically with
$s$. 
For this   reason we consider the calculation of $< |S|^2 >$ in a 
multi channel eikonal model, to be   an important check of the validity of
this
approach to  calculating  the  SC.

A hard LRG b-space amplitude, uncorrected by the survival probability, is
given in our formalism by 
\beq \label{4.5}
a_H\,\,\,=\,\,\a^2\,a^H_1\,\,+\,\,\b^2\,\,a^H_2
\eeq
where $a_1^H$ and $a_2^H$ correspond to processes initiated by our
$\Psi_1$ and $\Psi_2$ wave functions. As previously, we impose the
restriction that no inelastic rescattering modifies our hard process, by 
multiplying each component of $a_H$ by the corresponding $e^{-\Omega_n}$.  

The survival probability in our two channel  model is given by:
\beq \label{4.6}
< \mid S (\,s\,) \mid^2>\,\,\,=\,\,\,\frac{N(s)}{D(s)}\,\,,
\eeq
where 
\beq \label{4.7}
D(s)\,\,=\,\,\pi\,\int^{\infty}_0\,\,d b^2\,\,\left(\,\sum^2_{i =
1}\,\,a^H_i\,P_i(s,b)\,\right)\,\,=
\eeq
$$
\pi\,\int^{\infty}_0\,\,d b^2\,\,\left(\,\,e^{ -
\O_1(s,b)}\,\{\,1\,\,-\,\,\b^2\,[\,1\,-\,e^{- \Delta
\O(s,b)}\,]\,\}\,a^H_1\,\,+\,\, \b^2\,e^{ - \O_1(s,b)}\,e^{- \Delta 
\O(s,b)}\,a^H_{SD}\,\right)\,\,;
$$
and
\beq \label{4.8}
N(s)\,\,=\,\,\pi\,\int^{\infty}_0\,\,d
b^2\,\,\left(\,\sum^2_{i =   
1}\,\,a^H_i\,\right)\,\,=
\pi\,\int^{\infty}_0\,\,d
b^2\,\,\left(\,\,a^H_1\,\,+\,\,a^H_{SD}\,\right)\,\,.
\eeq

$< \mid S (\,s\,) \mid^2>$ can be calculated numerically once our free
parameters have been fixed provided we know $R^2_H$. We
have checked that taking $R^2_H = 8 \,GeV^{-2}$\cite{PSI}, we obtain $<
\mid S (\,s\,) \mid^2>$ values in the Tevatron range which are in a fair
agreement with the D0 data \cite{D0}. The calculated energy dependence of
$< \mid S (\,s\,) \mid^2>$ is given by 
$$
\frac{ < \mid S (\,\sqrt{s}\,=\,630\,GeV) \mid^2>}{ < \mid S
(\,\sqrt{s}\,=\,1800\,GeV) \mid^2>}\,\,=\,\,1.3\,-\,1.4
$$
to be compared with D0 value of $2.2 \,\pm\,0.8 $.
 In as much as this
is just a consistency check and not an actual
fit, we consider these results to be reasonable. A improved and detailed
analysis
of LRG events in a multi channel eikonal model will be published soon
\cite{GLMLAST}.

\section{Discussion}
The  following are the main features and predictions of our suggested
model.

\begin{enumerate}

\item\,\,\, The consistent treatment of the diffractive dissociation
channel in our model leads to a good description of the experimental data
on the single diffraction in ISR - Tevatron energy range. We reproduce the
energy dependence of the
ratio $\s_{SD}/\s_{el}$  as well as its value ( see Fig. 8a ) in 
agreement
with the experimental data.

 \item\,\,\, Our input Pomeron has $\Delta_P\,=\,0.126$ which
determines
our high energy predictions\,\,\, ( see Figs. 1 - 4 ). For LHC, $\sqrt{s}
\,=\,14\,TeV$, we predict $\s_{tot}\,=\,104.3\,mb$. Our prediction is
slightly higher than the DL  prediction of $101.5 \,mb$
and is compatible with a recent prediction of Block et al.\cite{BLOCK}
of $108\,\pm\,3.4\,mb$. The model suggested in Ref. \cite{BLOCK} is a
single channel eikonal calculation with a Regge input. DL
 approach is a
non screened Regge model corrected by a weak
Pomeron  -  Pomeron cut. The small differences between the  $\s_{tot}$
predictions at LHC energies of these rather different models, re-enforces
our observation
that SC are weak  for $\s_{tot}$.

\item\,\,\,Our LHC predictions for the elastic channel are
$\s_{el}\,=\,24.5\,mb$ and $B_{el}\,=\,20.5\,GeV^{-2}$. Our $\s_{el}$
prediction is about 15\% lower than the prediction of Ref. \cite{BLOCK}.
This is compatible with the small difference between the two $\s_{tot}$
predictions,     combined with our predicted value of $B_{el}$ being
higher
than the prediction of Block et al.

\item\,\,\, Some insight into the difference between a non screened  DL
type $b$-space elastic amplitude, and our screened $a_{el}(s,b)$ is
provided Fig.6. As seen in Fig.6a the DL non screened amplitude  violates
$s$-channel unitarity ( to be distinguished from the Froissart bound 
) at small impact parameters starting from  $\sqrt{s}\,\simeq\, 3\,TeV$.
The introduction of a weak P-P cut \cite{DL} reduces $a^{DL}_{el}(s,b)$ 
at small $b$ by approximately 10\%, leaving the DL prediction for
$\s_{tot}$ practically unchanged. In comparison, our screened
$a_{el}(s,b)$ is considerably lower and wider ( see Fig.6b ).
This behaviour is a consequence of both screening and the existence of
a competing diffractive channel.

\item\,\,\, As we have noted, the investigation of $\s_{SD}$ in the
ISR-Tevatron energy range provides strong support for the importance of SC
at existing collider energies. It is interesting to compare our approach
with the Pomeron flux renormalization suggested by Goulianos  \cite{DINO}. 
In the eikonal model, the  damping due to
unitarity  in 
the diffractive channel is given by   $e^{-\O_i(s,b)}$, where $i$ denotes
the rescattering  channel  considered. As such, the damping is
$b$-dependent,
with maximum  suppression at small $b$    resulting from the fact that the
input $\O_i(s,b)$ are central. As we saw, the damping  is smooth and
becomes  significant at ISR energies. In Ref.\cite{DINO}, a unitarity flux 
correction is
applied only when  the Pomeron flux exceeds unity. This amelioration
 is $b$-independent. Numerically, our $\s_{SD}$ has a stronger energy
dependence than  suggested in Ref. \cite{DINO}. In the ISR-Tevatron
range the difference between the predictions is small. However, at LHC,
$\sqrt{s}\,=\,14\,TeV$, we predict $\s_{SD}\,=\,12\,mb$ to be compared
with Goulianos' estimate  of $10.5\,mb$. 

\item\,\,\, A simple Gaussian input for the opacity leads to dips in
$\frac{d\sigma_{el}}{dt}$ at lower values of t than in the data. To obtain
these dips at their observed experimental values, the Gaussian form of
$\Omega(s,b)$ has to be replaced by a dipole form\cite{DIP}. We consider
this to be a relatively minor technical modification of the calculation.
 We note that the dip structure can be
reproduced in an eikonal calculation\cite{DIP}, as well as by the
introduction of a weak P-P cut\cite{DL}, leaving the DL results on
$\sigma_{tot}$ and $\sigma_{el}$ virtually unchanged.

\item \,\,\, The distinctive features of our diffractive $b$-space
amplitude are visible in Fig.7. Our $a_{SD}(s,b)$ is peripheral. This is
to be compared with Goulianos et al. \cite{DINO} who produce a SD
amplitude proportional to the unscreened DL amplitude shown in Fig.6a.
Our model is, thus, compatible with the Pumplin's bound \cite{PUM}

\beq \label{5.1}
| a_{D}(s,b)|^2\,\,=\,\,| a_{el}(s,b)|^2\,\,\,+\,\,\,| a_{SD}(s,b)
|^2\,\,\,\leq\,\,\,1\,\,,
\eeq
which translates, after a $b$-integration, to 
\beq \label{5.2}
\s_{el}(s)\,\,\,\,+\,\,\,\,\s_{SD}(s)\,\,\,\leq\,\,\,\frac{1}{2}\,\s_{tot}(s)
\,\,.
\eeq
In Fig.8 we show the energy dependence of
$R_{el}\,=\,\frac{\s_{el}}{\s_{tot}}$,
$R_{SD}\,=\,\frac{\s_{SD}}{\s_{tot}}$ and $R_D\,=\,R_{el} \,+\,R_{SD}$
( Fig.8a ),  and the energy dependence $| a_D(s,b) |^2\,\,=
\,\,|a_{el}(s,b)|^2\,\,+\,\,| a_{SD}(s,b)|^2$ at $b$ =0 ( Fig.8b ). All
ratios  are well below the Pumplin bound. We have
checked that the bound is saturated at the non realistic ultra high
energies of $\sqrt{s}\,\simeq\,10^9 \,GeV$. The  Goulianos
model differs in as much as it violates the Pumplin  bound at small
$b$.   This is a consequence of both $a_{el}(s,b)$ and $a_{SD}(s,b)$
peaking at $b$ = 0. We note that the Pumplin's bound was proven in a
multichannel eikonal model. Its validity in other screening
models is not clear \cite{KAID}. A peripheral $b$-space structure for
$a_{SD}$ implies dips or breaks in $\frac{d \s_{SD}}{d t}$ at small $t$,
where the details depend on form of the input  taken in the eikonal
calculation. Some indications for such a structure have been observed in
ISR  SD experiment with a very small diffractive mass \cite{DIPS}. 
More accurate experimental data are required to investigate this aspect
further.

\item\,\,\, A more detailed treatment  of $< \mid S\mid^2 >$ in a
multichannel
eikonal model is in preparation \cite{GLMLAST}.  In the present  paper we
have
calculated this observable as a  check of   our model. We
obtain a reasonable energy dependence  of  $< \mid S\mid^2 >$ even though
$R_D$ is
almost energy independent in the ISR- Tevatron energy range ( see Fig.8a
).  
The resulting energy dependence of $< \mid S\mid^2 >$ in our model is not
surprising as each channel is suppresed by $e^{-\O_i(s,b)}$. As long as
the SC  are small, the energy dependence of 
$< \mid S\mid^2 >$ is determined by $R_D$ which is almost flat in the
energy range of interest. When the SC become significant, the relevant
ratios are $R_{el}$ and $R_{SD}$, both of which show sufficient energy
dependence to induce a final $< \mid S\mid^2 >$ which is compatible with
the D0 measurements \cite{D0}.

\item\,\,\, In a recent preprint Cudell et al.\cite{CUD} come to a 
conclusion similar to ours  for $\s_{tot}$. However, Cudell et al. use
this result to
question whether unitarity corrections are necessary at all. We wish to
emphasize again that the value  of SC differs for different channels.
They are mild for the elastic channels but appreciable for $\s_{SD}$ and 
$< \mid S\mid^2 >$. Since we believe that the same colour singlet (
Pomeron
) mediates all these reactions, we advocate not  drawing  general
conclusions
which are based on the finding in one channel only. 

 \end{enumerate}

\begin{figure}
\begin{tabular}{c c}
\psfig{file= 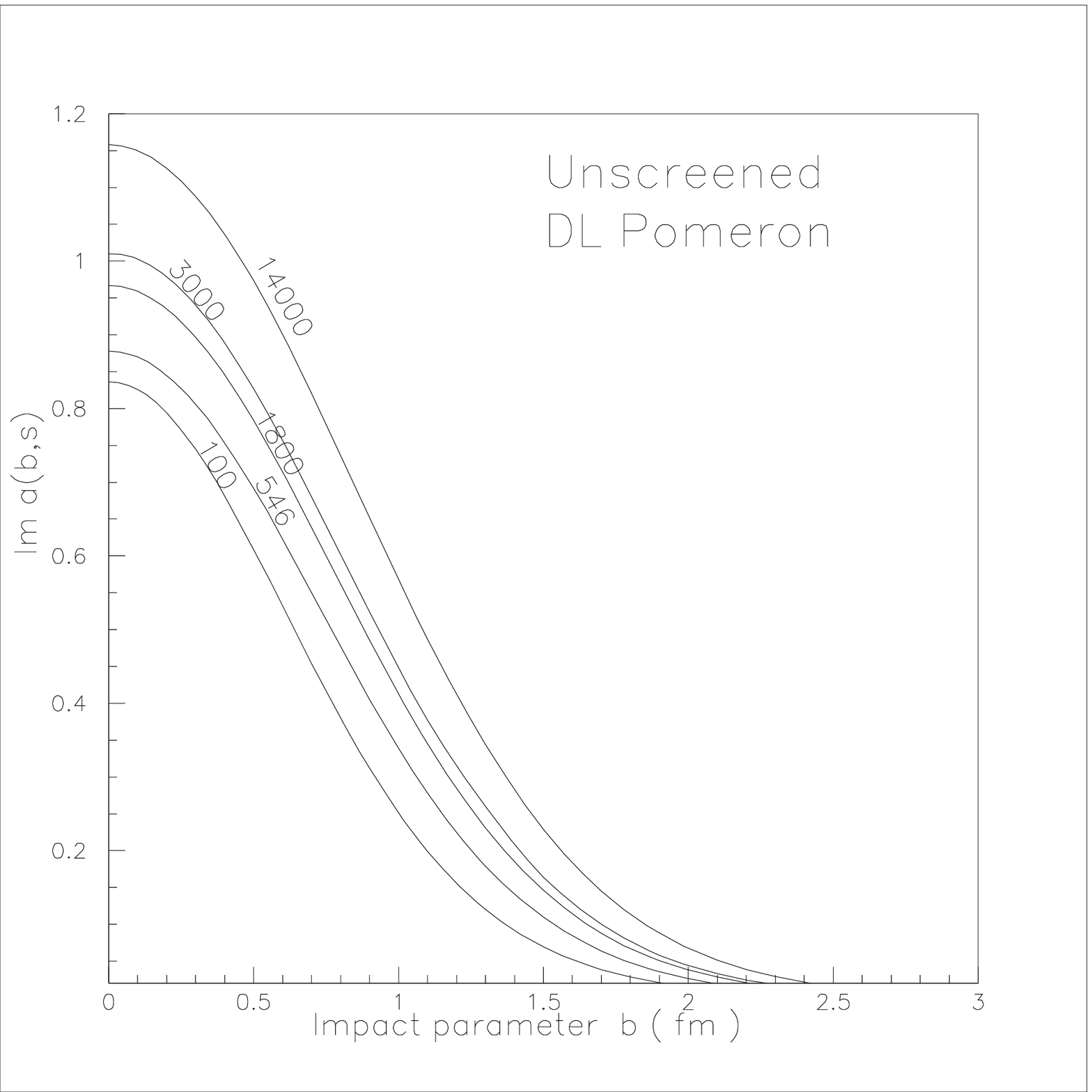, width=80mm,height=70mm} &\psfig{file=
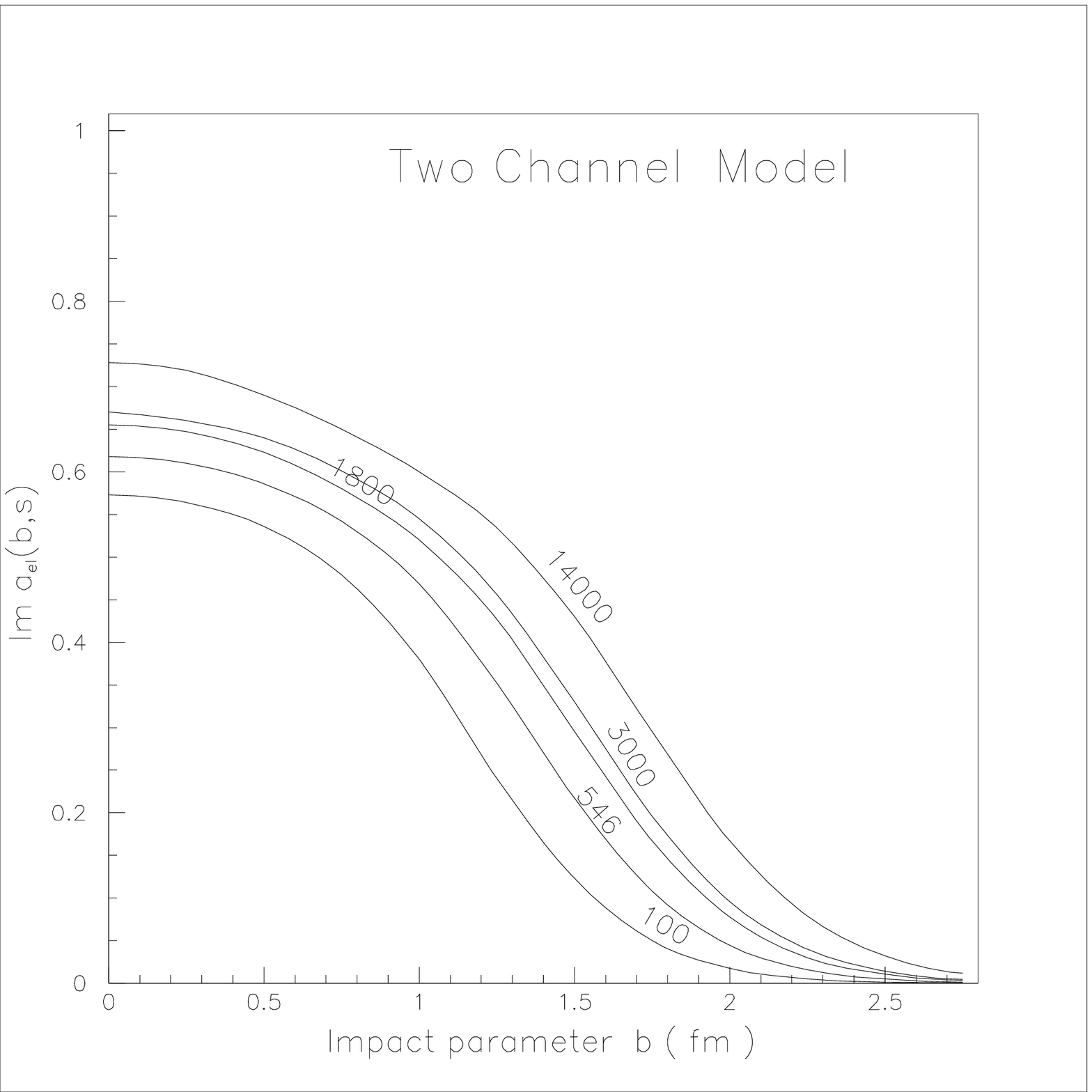,width=80mm,height=70mm}\\
Fig. 6-a & Fig. 6-b\\
\end{tabular}
\vspace{2cm}
\caption{ \it The $b$ - dependence of the elastic amplitude for the
unscreened DL  Pomeron ( Fig.6a )   and in our two channel  model (
Fig. 6b ).}
\label{fig6 }
\end{figure}

\begin{figure}
\centerline{\psfig{file= 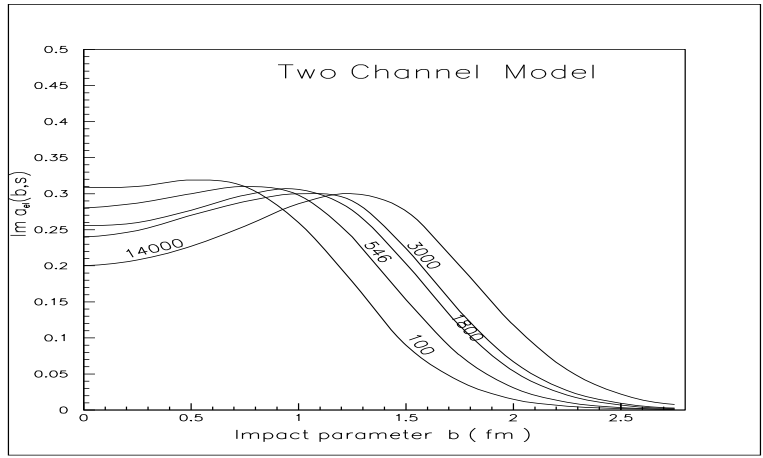,width=100mm,height=60mm}}
\caption{ \it The $b$ - dependence of the single diffraction  amplitude
 in two channel  model.}  
\label{fig7}
\end{figure}

\begin{figure}
\begin{tabular}{c c}
\psfig{file= 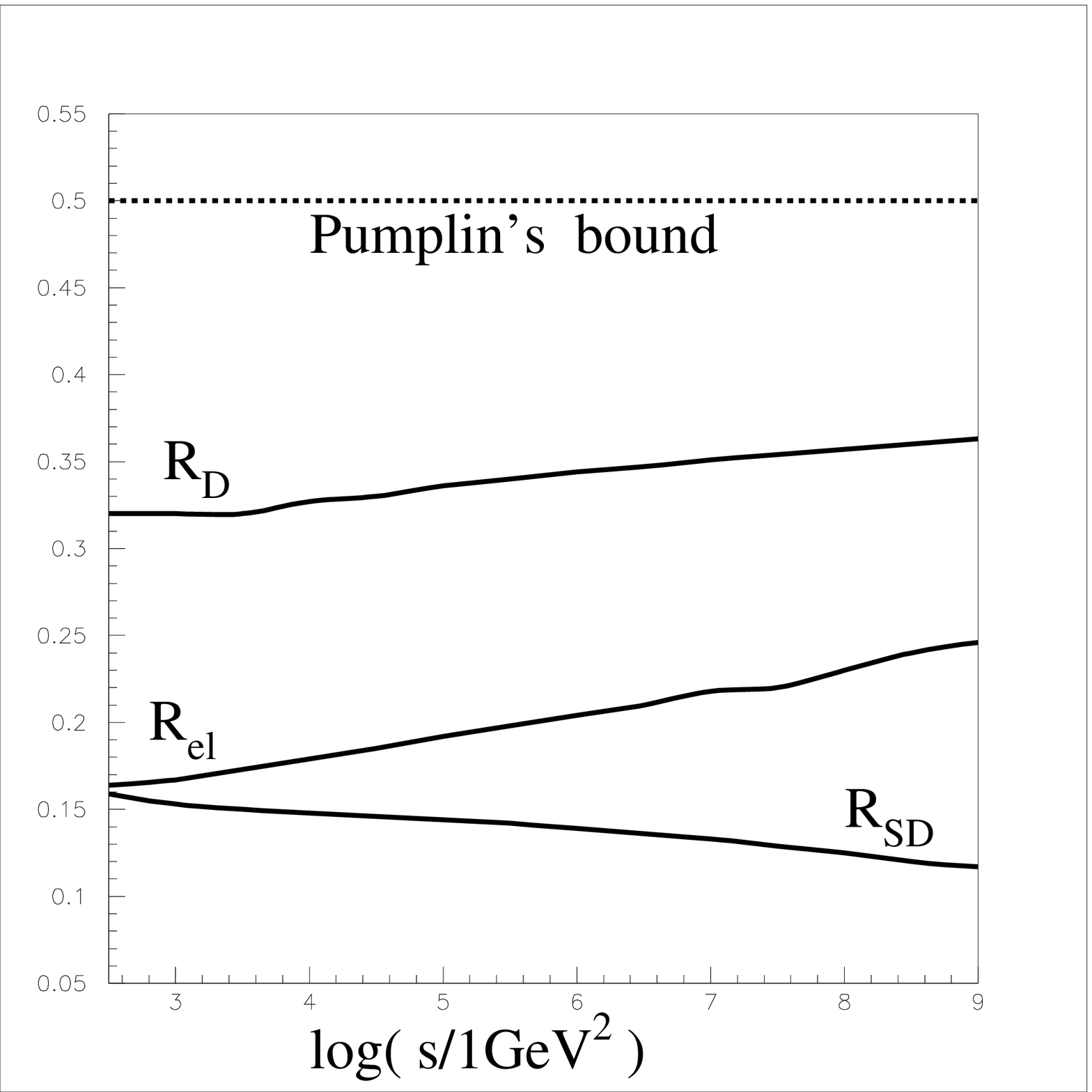, width=80mm,height=70mm} &\psfig{file=
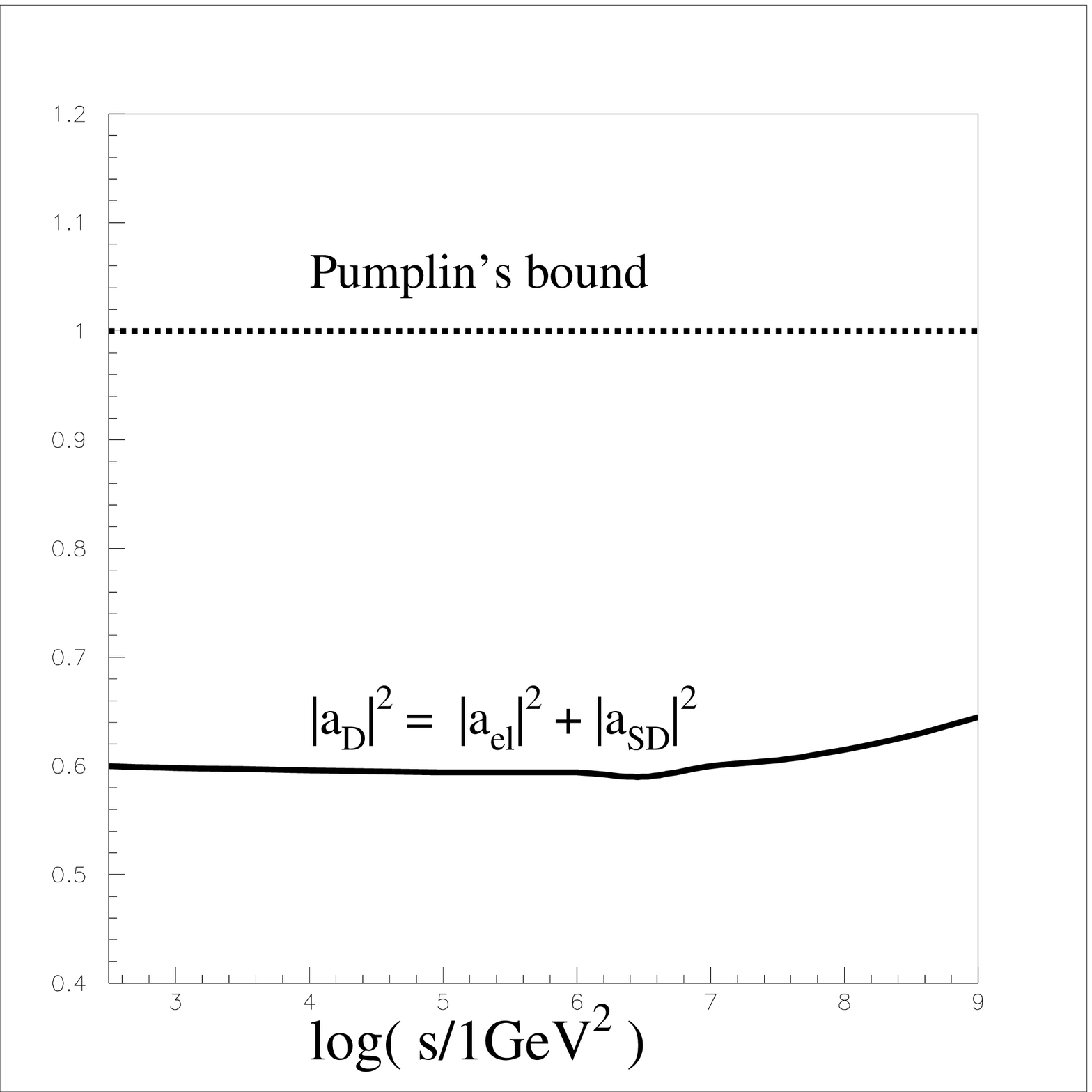,width=80mm,height=70mm}\\
Fig. 8-a & Fig. 8-b\\
\end{tabular}
\vspace{2cm}
\caption{ \it  Energy behaviour of the ratios $R_D\,=\,( \sigma_{el} +
\sigma_{SD})/\sigma_{tot}$, $R_{el}\,= \,\sigma_{el}/\sigma_{tot}$ and
$R_{SD}\,=\,\sigma_{SD}/\sigma_{tot}$  (Fig.8a ) and $
|a_{D}(s,b=0)|^2\,=\,|a_{el}(s,b=0)|^2 \,+\,| a_{SD}(s,b=0)|^2$  (
Fig.8b).
The dotted line in both figures shows the Pumplin bound.}
\label{fig8 }  
\end{figure}

{\bf Acknowledgements:} This research was supported in part by the Israel
Science Foundation, founded by Israel Academy of Science and Humanities.

\end{document}